# Interplay between electronic dephasing and localization in finite-sized Chern insulator

Yunhe Bai[1,10]†, Yuanzhao Li[1,2]†, Jianli Luan[1,2]†, Yang Chen[1], Zongwei Gao[2], Wenyu Song[1], Yitian Tong[1], Jinsong Zhang[1,3,4], Yayu Wang[1,3,4], Junjie Qi[2], Chui-Zhen Chen[5,6], Hua Jiang[7], X. C. Xie[4,7,8], Ke He[1,2,3,4]*, Yang Feng[2]*, Xiao Feng[1,2,3,4]*, and Qi-Kun Xue[1,2,3,4,9]

[1]State Key Laboratory of Low Dimensional Quantum Physics, Department of Physics, Tsinghua University, Beijing 100084, China;
[2]Beijing Academy of Quantum Information Sciences, Beijing 100193, China;
[3]Frontier Science Center for Quantum Information, Beijing 100084, China;
[4]Hefei National Laboratory, Hefei 230088, China;
[5]School of Physical Science and Technology, Soochow University, Suzhou 215006, China;
[6]Institute for Advanced Study, Soochow University, Suzhou 215006, China;
[7]Institute for Nanoelectronic Devices and Quantum Computing, Fudan University, Shanghai 200438, China;
[8]International Center for Quantum Materials, School of Physics, Peking University, Beijing 100871, China;
[9]Southern University of Science and Technology, Shenzhen 518055, China;
[10]Present address: Pritzker School of Molecular Engineering, The University of Chicago, Chicago, IL, USA.
†*These authors contributed equally to this work.*
*Corresponding author*. E-mail: kehe@tsinghua.edu.cn (K. H.); fengyang@baqis.ac.cn (Y. F.); xiaofeng@mail.tsinghua.edu.cn (X. F.)

## Abstract

Anderson localization is anticipated to play a pivotal role in the manifestation of the quantum anomalous Hall effect, akin to its role in conventional quantum Hall effects. The significance of Anderson localization is particularly pronounced in elucidating the reasons behind the fragility of the observed quantum anomalous Hall state in the intrinsic magnetic topological insulator $MnBi_2Te_4$ with a large predicted magnetic gap. Here, employing varying sized $MnBi_2Te_4$ micro/nano-structures fabricated from a single molecular-beam-epitaxy-grown thin film, we have carried out a systematic size- and temperature-dependent study on the transport properties of the films regarding the quantum anomalous Hall states. The low-temperature transport properties of the finite-sized $MnBi_2Te_4$ samples can be quantitatively understood through Anderson localization, which plays an indispensable role in stabilizing the ground states. At higher temperatures, the failure of electron localization induced by an excessively short electronic dephasing length is identified as the cause of deviation from quantization. The work reveals that electronic dephasing and localization are non-negligible factors in designing high-temperature quantum anomalous Hall systems.

**Main**

The quantum Hall (QH) effect and Anderson localization have a close relationship[1]. Anderson localization is crucial for the robust quantized plateaus of Hall resistivity characterizing the QH effect. On the other hand, QH systems provide a good platform to quantitatively study Anderson localization in the two-dimensional (2D) case. The conductivity of a 2D electron system subject to inevitable disorder, according to Anderson localization[2,3], decreases exponentially with the sample size $L_S$: $\sigma_{xx} \sim \exp(-L_S/\xi)$ where $\xi$ is the localization length. Since Anderson localization results from the interference of electrons, the electronic dephasing length $l_\phi$, which characterizes the length scale of quantum-interference phenomena, acts as the effective sample size when it is smaller than $L_S$ (ref. 4). The critical behaviors related to Anderson localization have been investigated in QH systems through temperature-dependent transport properties around the quantum phase transitions (QPTs) between QH plateaus[5], in which temperature is used to tune $l_\phi$ relative to $\xi$ and $L_S$. Nevertheless, there were very few experiments directly studying Anderson localization at quantized plateaus (far from QPTs), because $\xi$ in plateau states is too small (several tens of nanometers, on the order of the magnetic length[6]) for $l_\phi$ or achievable $L_S$. Such studies, which can directly unveil the influences of Anderson localization on the quantized transport properties of chiral edge states, are highly anticipated for understanding a special QH effect—the quantum anomalous Hall (QAH) effect.

The QAH effect is a variant of QH effect contributed by magnetization-induced topologically non-trivial bands instead of Landau levels[7]. The effect provides a road towards widespread electronic applications of dissipationless QH edge states, which is, however, obstructed by the rather low temperature needed to achieve it (below 1 K in most cases)[8–12]. The recently discovered intrinsic magnetic topological insulator (TI) MnBi$_2$Te$_4$ has a magnetically induced surface state gap up to ~50 meV according to first-principles calculations[13–15], in principle capable of supporting the QAH effect at a rather high temperature. But the QAH state experimentally achieved in MnBi$_2$Te$_4$ is much more fragile than theoretically expected[9,16–23]. Since Anderson localization plays a key role in the robustness of usual QH effects, its role in the QAH effect and its occurring temperature should also be thoroughly considered. The QAH phase (or Chern insulator phase) is expected in MnBi$_2$Te$_4$ thin films both in the antiferromagnetic (AFM) ground state without magnetic field (only in odd-layer films) and in the ferromagnetic (FM) configuration under high magnetic field (regardless of film thickness)[13–15]. Although in both cases, the QAH phase has been achieved, the former one is more difficult, reported by only two groups so far[9,24]. In most MnBi$_2$Te$_4$ samples, normal insulator behaviors are observed around zero magnetic field in both odd- and even-layer samples[9,17,19–22]. The QPT from a normal insulator to the QAH phase occurs with an increasing magnetic field, allowing us to investigate the critical behaviors. Besides, as shown below, $\xi$ in MnBi$_2$Te$_4$ is around 1 μm in the QAH phase, much longer than that in usual QH systems, which makes possible a size-dependent study with standard fabrication techniques. In this work, we systematically investigated finite-sized MnBi$_2$Te$_4$ thin films within the quantized regime with both the size- and temperature-dependent transport measurements, both near and far from QPT, which presents a comprehensive picture on the intricate relationship between Anderson localization and QAH state.

A considerable challenge for the study is the rather low repeatability of MnBi$_2$Te$_4$ samples with quantized transport properties (even those only quantized at high magnetic field). In this situation, the only feasible way for a size-dependent investigation is by measuring a series of devices with different sizes fabricated from one homogeneously quantized MnBi$_2$Te$_4$ epitaxial film (thin flake samples are too small for the task). Homogeneous MnBi$_2$Te$_4$ epitaxial films exhibiting quantized anomalous

Hall resistivity in the FM configuration have been achieved in our previous work[23]. For the present study, we developed a set of fabrication procedures based on electron-beam lithography (EBL) that enables us to make structures down to several hundred nanometers from $MnBi_2Te_4$ epitaxial thin films without destroying their quantized transport properties (see Methods for the details of sample preparation). Figure 1a shows an optical image of the samples studied in this work: including eight Hall bar devices (two of them failed) with the widths ($W$) ranging from 0.3 μm to 30 μm, each with an individual top gate, fabricated on a 5-SL $MnBi_2Te_4$ film. The cyan box in Fig. 1a indicates one of the devices, and the zoom-in image of its Hall bar region is shown in Fig. 1b. The aspect ratio is 1 for devices with $W \geq 5$ μm and 2 for those with $W < 5$ μm (see Methods for more details). Hereinafter, $W$ and length ($L$), as defined in Fig. 1b, are used to represent device dimensions.

Figures 1c and 1d display the magnetic field ($H$) dependences of the Hall resistivity ($\rho_{yx}$) and the longitudinal resistivity ($\rho_{xx}$) of the largest device ($W = 30$ μm), respectively. The gate voltage ($V_g$) is tuned to the charge-neutral point (CNP). At high magnetic field, say 9 T (FM configuration), $\rho_{yx}$ is perfectly quantized, and $\rho_{xx}$ drops to ~0.002 $h/e^2$ at 20 mK ($h$ is the Planck constant, and $e$ is the elementary charge), clearly indicating a QAH state (Chern number $C = -1$). It is the first time that full quantization of $\rho_{yx}$ with vanishing $\rho_{xx}$ has been achieved in $MnBi_2Te_4$ epitaxial thin films (under high magnetic field). The device maintains nearly full quantization up to the temperature $T = 0.7$ K. Around zero magnetic field (AFM configuration), $\rho_{xx}$ increases abruptly with decreasing temperature, up to ~10 $h/e^2$ at $T = 20$ mK, and $\rho_{yx}$ shows a hysteresis loop with the zero-field value far below $h/e^2$. The $\rho_{yx}$–$\mu_0 H$ curves become noisier at lower temperatures, mainly due to large device resistance. With increasing magnetic field, $\rho_{xx}$–$\mu_0 H$ curves at different temperatures cross at one point ($H_{cr}$ in Fig. 1d), suggesting a QPT occurring here. These observations, together with the flow diagram to be discussed later (Fig. 3i), suggest that the film around zero magnetic field is a normal insulator ($C = 0$)[25].

To assess the homogeneity of the film, we present the magnetic ordering temperature ($T_N$), coercive field ($H_c$), and $H_{cr}$ of the successful devices of different sizes in Fig. 1e, extracted from Fig. 2 ($H_{cr}$ of $W$ = 0.3 μm device is excluded due to too noisy $\rho_{xx}$). All these quantities exhibit minimal variation across different devices. Figure 1f shows the $1/T$-dependent longitudinal conductivity ($\sigma_{xx}$) of the devices (except for the $W$ = 0.3 μm one because of significant noise during $T$ variations) at -9 T. The linear relation observed in all the curves at high-temperature region demonstrates a typical thermal activation behavior, with the slope determined by the activation gap $\Delta$. All the curves have a similar slope, and the fitted $\Delta$ of the devices (inset of Fig. 1f) are all around 2 K. The low-temperature part of the curves in Fig. 1f is size-dependent and will be discussed below and in Fig. S2. Obviously, the magnetic and electronic properties of the different devices on the chip are uniform enough for a size-dependent study.

Figure 2 displays the transport properties of the six devices with different widths. The magnetic field-dependent $\rho_{yx}$ and $\rho_{xx}$ ($V_g$ is set at CNP) at different temperatures (from 0.02 K to 3 K) are shown in Figs. 2a and 2b, respectively. The $V_g$-dependent $\rho_{yx}$ (blue) and $\rho_{xx}$ (red) at $T$ = 0.02 K and $\mu_0 H$ = -9 T are depicted in Fig. 2c. All the devices exhibit similar behaviors to the $W$ = 30 μm device, indicating the QAH phase at high field and the normal insulator phase at low field. In the $V_g$-dependent curves in Fig. 2c, $\rho_{yx}$ manifests a perfectly quantized plateau (blue curves) in all the devices except for the smallest one ($W$ = 0.3 μm). Even in this device, the $\rho_{yx}$ maximum only decreases to ~0.95 $h/e^2$. On the other hand, $\rho_{xx}$ obviously deviates from zero in the $W$ = 10 μm device and increases gradually with decreasing $W$.

A straightforward explanation for the deviation from quantization with reduced $W$ could be the direct tunneling between the chiral edge states of opposite edges. However, this scenario implies that the width of the chiral edge states is several microns, too large compared with that reported in magnetically doped TIs[26,27]. Moreover, the tunneling between two chiral edge states is expected to increase exponentially with decreasing $W$ (ref. 7). Yet, even though $\rho_{xx}$ becomes non-vanishing from $W$ = 10 μm, the deviation from quantization is not significant in the $W$ = 0.3 μm device ($\rho_{yx}$ ~ 0.95 $h/e^2$; $\rho_{xx}$ ~ 0.1 $h/e^2$), which means that chiral edge states still dominate the transport properties. Clearly, our observations cannot be simply interpreted by the direct scattering between the chiral edge states[7].

To understand the observation, we summarize $\rho_{yx}$ and $\rho_{xx}$ as functions of $W$ at different temperatures with $\mu_0 H$ = -9 T (0 T) in Figs. 3a (3d) and 3b (3e), respectively. In Fig. 3a ($\rho_{yx}$ (-9 T)), at the lowest temperature (blue), $\rho_{yx}$ remains quantized in all devices except for the $W$ = 0.3 μm one; at higher temperatures, $\rho_{yx}$–$W$ curves show a maximum at $W$ = 2 μm, which will be discussed later. In Fig. 3d ($\rho_{yx}$ (0 T)), only the $W$ dependences of $\rho_{yx}$ at 2 K and 3 K are displayed, because the data measured at lower temperatures are too noisy to be read (see Fig. 2a). In Fig. 3b ($\rho_{xx}$ (-9 T)), at the lowest temperature (blue), $\rho_{xx}$ decreases with increasing $W$. Higher temperatures not only enhance $\rho_{xx}$ but also change its $W$ dependence. At 3 K (red), $\rho_{xx}$ exhibits an increasing trend with increasing $W$. In Fig. 3e ($\rho_{xx}$ (0 T)), at the lowest temperature (blue), $\rho_{xx}$ largely increases with increasing $W$. With increasing temperature, $\rho_{xx}$ decreases in all devices but more significantly in larger ones. As a result, at 3 K (red), $\rho_{xx}$ only slightly grows with $W$.

A similar trend shared by the $W$ dependences of $\rho_{xx}$ at high field (Fig. 3b) and zero field (Fig. 3e) is that in larger devices $\rho_{xx}$ exhibits stronger temperature dependences (it can be directly observed in Fig. 2b). The evident correlation between the temperature and size dependences of $\rho_{xx}$ is a signature of Anderson localization. The electron dephasing length $l_\phi$ increases with decreasing $T$ according to the relation $l_\phi \propto T^{-p/2}$ where $p$ is the inelastic-scattering-length exponent[1,4], and meanwhile, localization length $\xi$ is insensitive to $T$. Therefore, at sufficiently low $T$, $l_\phi$ can exceed the actual device size $L_S$, and the sample conductivity $\sigma_{xx}$ will decrease with increasing $L_S$ due to Anderson localization. Schematics illustrating the scenarios for small and large devices at specific low $T$ are depicted in Figs. 3g and 3h, respectively. At higher temperatures such that $l_\phi$ becomes smaller than the actual device size, the sample conductivity will be insensitive to $L_S$ but sensitive to $l_\phi$ and $T$.

Figures 3c and 3f display the $W$ dependences of $\sigma_{xx}$, derived from the $\rho_{yx}$ and $\rho_{xx}$ data, at different temperatures with $\mu_0 H$ = -9 T and 0 T, respectively. At the lowest temperature (blue), $\sigma_{xx}$ decreases with increasing $W$, at both high and zero magnetic fields, qualitatively consistent with Anderson localization. The electron localization drives the film towards the insulating ground states, either normal ($C = 0$) or Chern insulator ($C = -1$). In Fig. 3i, the ($\sigma_{xy}$, $\sigma_{xx}$) data of different device sizes (represented by different colors) and different magnetic fields are plotted (flow diagram). The data points at 0 T and -6 T are denoted by hollow triangles, while others are depicted with solid lines (the data of -6 T instead of -9 T are shown because the latter ones are all around the stable point). With increasing device size, the ($\sigma_{xy}$, $\sigma_{xx}$) data points at 0 T and -6 T flow towards the stable fixed points (0, 0) and ($e^2/h$, 0), respectively. This observation is well-consistent with the finite-size scaling behavior in QH/QAH systems due to Anderson localization, a phenomenon well-established in theory[1,28] but scarcely demonstrated directly in experiments in usual QH systems.

To confirm the role of Anderson localization quantitatively, we try to extract $\xi$ and $l_\phi$ values from the experiment. Figure 4a shows the current-dependent $\rho_{xx}$ (-9 T) for each device at 20 mK, obtained using a tunable dc bias current $I$ in parallel with a small ac excitation current of 1 nA. When $I$ exceeds a critical value $I_c$, $\rho_{xx}$ significantly increases, due to the breakdown of the QAH phase. Within the variable-range hopping regime, in-gap states are accelerated by a transverse electric field ($E_y$) within a typical length $\xi$ (ref. 6,29,30), and the effective electron temperature ($T_{eff}$) is elevated due to dissipation of the work done by the electric field force: $k_B T_{eff} \sim eE_y \cdot \xi/2 = e\rho_{yx}I/W \cdot \xi/2$. $\xi$ values are estimated by fitting the relation (Fig. 4b), and the fitted $\xi$ ranges from 0.4 to 1.2 μm in different devices (inset of Fig. 4c). The renormalized localization length $\xi/W$ decreases with increasing $W$ (Fig. 4c), quantitatively consistent with the numerical calculation results in 2D systems in Anderson localization[31,32].

The electronic dephasing length $l_\phi$ can be obtained by analyzing the critical behaviors near the QPT between the $C = 0$ state at zero magnetic field and the $C = -1$ state at high magnetic field[1,33]. According to the finite-size scaling theory[1,33], the maximum slope $|\partial\sigma_{xy}/\partial\mu_0 H|$ (inset of Fig. 4d) around $H_{cr}$ is proportional to $T^{-\kappa}$, where $\kappa$ is the temperature scaling exponent. As the double-logarithmic plot of $\partial\sigma_{xy}/\partial\mu_0 H$ with $T$ shown in Fig. 4d, the data of different devices coincide on a single line above approximately 1 K with the slope $\kappa \sim 0.49 \pm 0.02$ and gradually saturate at different values at low temperatures. The behavior can be experimentally expressed as[34]:

$$\partial\sigma_{xy}/\partial\mu_0 H \sim (T^p + l_0^2/L \cdot W)^{-1/2\nu}, \qquad (1)$$

where $l_0$ is a prefactor linked to $l_\phi$, and $\nu = p/2\kappa$ is the localization-length exponent. The first and second terms in the parentheses represent the scattering rates from inelastic processes and sample boundaries, respectively[34]. At sufficiently low temperature, inelastic scattering (the first term) is suppressed, and we can estimate the critical exponent $\nu \sim 2.56 \pm 0.05$ by fitting the data with the second term, as shown in Fig. 4e. $p \sim 2.49 \pm 0.14$ is obtained from the relation between critical exponents, and the fitting curves well capture the features as shown in Fig. 4d. The experimentally extracted critical exponent $\nu$ is close to the value expected for a phase transition in the universality class of QH/QAH effect[5,33] and previous experiments[35–37], supporting the validity of the above analyses. The temperatures at which the slope drops to the saturated values ($T_s$) are obtained by the intersection of the two straight asymptotes fitted to the two parts of the curves in Fig. 4d. Since $T_s$ corresponds to the temperature at which $l_\phi$ roughly equals $W$, $l_\phi$ at different temperatures are obtained, as shown in Fig. 4f. The estimated $l_\phi$ at 2 K in our study is of the same magnitude as those obtained in $Bi_2Se_3$-family TIs through weak anti-localization, typically around hundreds of nanometers[38,39].

The estimated $\xi$ and $l_\phi$ values provide quantitative evidences for the contribution of Anderson localization to the size dependence of $\sigma_{xx}$ (-9 T) at 20 mK. At this temperature, $\xi$ is about 1 μm, and $l_\phi$ exceeds several tens of μm (see Figs. 4c and 4f). According to Anderson localization, $\sigma_{xx}$ is expected to show an obvious decrease as the device size increases from $\xi$ to several times $\xi$, just as the observation. The $\sigma_{xx}$ (-9 T)–$W$ data at 20 mK can be well-fitted with the relation $\sigma_{xx} \propto \cosh^{-2}(W/\xi)$ obtained by the scattering matrix theory (the pink dashed line in Fig. 3c) with $\xi = 1$ μm (ref. 40), which can be translated to an exponential decay in the context of large device sizes, as indicated by Anderson localization (Fig. 3h). Meanwhile, the presence of in-gap states contributes to enhanced scattering between edge states at opposite edges in small device sizes comparable to $\xi$, facilitated by resonant tunneling (Fig. 3g)[41].

As $l_\phi$ decreases with increasing temperature down to below the device size, $\sigma_{xx}$ should become insensitive to $W$. From Fig. 3c, the size dependence of $\sigma_{xx}$ indeed becomes less significant as $T$ increases up to 0.7 K. However, at higher temperatures, an obvious increase in $\sigma_{xx}$ with increasing device size is distinguished. This observation cannot be understood solely with Anderson localization. We notice that above 1 K, $l_\phi$ is reduced below $\xi$ (see Figs. 4c and 4f), which means that bulk electrons cannot be localized but contribute to conductivity. According to classical Ohm's law, conductivity should be invariant with device size, in contrast to the current observation. The solution to this puzzle may lie in the fact that in the smallest device ($W$ = 0.3 μm), $l_\phi$ is close to the device size, and electrons basically maintain phase coherence throughout the entire device. In larger devices, the device size (especially $L$) relative to $l_\phi$ increases, and consequently, electrons in chiral edge states experience more phase-broken inelastic scatterings into dissipative bulk bands[42], which enhances $\sigma_{xx}$. Actually, only considering the $\sigma_{xx}$ of bulk electrons may also increase with the sample size when it is near $l_\phi$ due to phase-broken scatterings in certain situations[43]. This understanding can also explain the $\rho_{yx}$ (-9 T)–$W$ curves in Fig. 3a where $\rho_{yx}$ at $W$ = 2 μm decreases more slowly with increasing temperatures than larger and smaller devices. In smaller devices, the more rapid decrease of $\rho_{yx}$ with increasing temperature can be attributed to enhanced scattering between two chiral edge states by elevated temperatures[41]. Meanwhile in larger devices, the more rapid $\rho_{yx}$ drop can be attributed to the enhanced $\sigma_{xx}$ due to electronic dephasing.

At last, we attempt to understand the data at zero magnetic field. At 20 mK, $\sigma_{xx}$ (0 T) also decreases with increasing device size, but more slowly compared to that at high field and saturates at 0.1 $e^2/h$ when $W$ exceeds 10 μm. Such behavior cannot be well-fitted by the theoretical relation of Anderson localization. The observation suggests that $\xi$ is larger at zero magnetic field, and there might exist conducting channels that cannot be localized. At higher temperatures, $W$ dependences of $\sigma_{xx}$ (0 T) are analogous to that at 20 mK (except for a slight increase in $\sigma_{xx}$ in the $W$ = 30 μm device), which implies that the system is little influenced by electron dephasing.

So, the observed temperature- and size-dependent transport properties of MnBi$_2$Te$_4$ epitaxial films, especially in the QAH regime under high magnetic field, can be well understood through the interplay of electron localization and dephasing. The large localization length (~1 μm) allows us to vary $l_\phi$ and the device size from below to above it, giving a relatively complete picture of the relationship between Anderson localization, electronic dephasing, and the QAH effect. Interestingly, in scanning microwave impedance microscopy (SMIM) studies of MnBi$_2$Te$_4$, the observed edge states also have a width of ~1 μm (ref. 22), in the same order of magnitude as the scale width of QAH edge states[44]. It is likely that SMIM detects signals not only from the chiral edge states but also from the bulk states within the localization length from the chiral edge states. The exact mechanism for this requires further investigation.

On the other hand, our results also demonstrate that electronic dephasing can be a serious obstacle to achieving a high-temperature QAH effect, a factor that has rarely been discussed before. Even for a QAH system with a large magnetic gap and a high magnetic ordering temperature, it cannot exhibit the QAH effect at a temperature at which $l_\phi$ is shorter than $\xi$. Conventional QH systems rarely encounter the problem since $\xi$ is typically only around several tens of nanometers, which is usually much smaller than $l_\phi$. In our epitaxial MnBi$_2$Te$_4$ films, $\xi$ is similar to that in Cr-doped (Bi,Sb)$_2$Te$_3$. It is understandable that the temperature to observe the QAH effect in the former does not significantly exceed that in the latter. We still cannot figure out what leads to the large $\xi$ in magnetic TIs. A recent magnetic force microscopy (MFM) study on our epitaxial MnBi$_2$Te$_4$ films unveils μm-scaled inhomogeneity of magnetic force signals[45], which may have some relationship with the μm-scaled $\xi$. Charge puddles induced by impurities and disorder may also enhance $\xi$.

The present size-dependent study not only unveils the important roles of electronic localization and dephasing in the QAH effect, which may inspire new insights for the exploration of high-temperature QAH effect, but also represents a progress towards the scalable fabrication of QAH micro/nano-structures based on the extremely fragile intrinsic magnetic TI MnBi$_2$Te$_4$. The techniques we developed can be applied to construct arrays of QAH devices for quantum resistance standards and QAH-based topological quantum computation.

# Methods

## MBE growth of MnBi$_2$Te$_4$ thin films

MnBi$_2$Te$_4$ thin films were epitaxially grown on sapphire (0001) substrates in a commercial MBE system (Omicron Lab10) with a base pressure lower than 2×10$^{-10}$ mbar. The sapphire substrates were annealed in a tube furnace at around 1100°C for 3 hours under a N$_2$:O$_2$ = 4:1 flow. Following this treatment, the sapphire substrate surfaces offered flat terraces and an optimal surface environment for the subsequent growth of MnBi$_2$Te$_4$ thin films. The treated substrates were then loaded into the MBE chamber and degassed at 400°C for 30 minutes before growth. High-purity Mn (99.9998%), Bi (99.9999%) and Te (99.9999%) were co-evaporated from Knudsen effusion cells. The substrate temperature was maintained at around 270°C during growth, and a 2-hour post-annealing under Te flux was applied at the same substrate temperature for better crystalline quality. In situ reflection high-energy electron diffraction (RHEED) was employed to monitor the sample quality. After the growth, the as-grown films were transferred to another MBE chamber and exposed to O$_2$ at a pressure of 1 × 10$^{-3}$ mbar for 1 hour at room temperature. An 8-nm-thick cadmium selenide layer was then deposited on top of the thin films as a capping layer to prevent uncontrolled oxidation, in the following fabrications and transport measurements.

## Device arrays fabrication

The 2 mm × 2 mm device arrays shown in the main text (Fig. 1a) were patterned on a single 3 mm × 5 mm chip for transport measurements. EBL was first employed to pattern the Hall bar electrodes and markers. Before depositing a 5 nm Ti/35 nm Au layer using electron beam evaporation, the insulating capping layer (cadmium selenide) was removed through etching to ensure better contact. The Hall bar devices, ranging from sub-micron to tens of microns in scale, were defined by a second-step EBL and then etched by argon ion milling. The aspect ratio ($L/W$, as defined in Fig. 1b) was set at 1 for devices with $W$ = 5 μm, 10 μm and 30 μm. The aspect ratio was constrained to 2 for devices with $W$ = 0.3 μm, 1 μm and 2 μm due to fabrication limitations. Subsequently, a 40-nm-thick AlO$_x$ layer was grown on top of the sample as a gate dielectric using atomic layer deposition. Finally, top-gate electrodes, composed of 5 nm Ti/ 25 nm Au, were deposited through a molybdenum mask.

**Electrical transport measurements**

Transport measurements were conducted using a Triton 450 dilution fridge with a base temperature of 20 mK produced by Oxford Instruments. The applied magnetic field in this study was perpendicular to the film plane. In ac measurements, the current was injected by a Keithley 6221 source meter or a lock-in amplifier SR830, and the voltage was detected by a SR830. In dc measurements, the current was injected by a Keithley 6221 and the voltage was detected by a Keithley 2182 nanovoltmeter in the delta mode. The top-gate voltage was applied using a Keithley 2400 multimeter. A 5 nA ac excitation current was predominantly applied in transport measurements for a good signal-to-noise ratio, unless otherwise specified. However, for the transport measurements of the 0.3-μm device at 20 mK (Figs. 2 and 3), a dc measurement was employed for reliable data. In the current-induced QAH breakdown measurements (Fig. 4a), a tunable dc bias current in parallel with 1 nA ac excitation current was applied using a Keithley 6221. All magneto-transport results were symmetrized for $\rho_{xx}$ data and anti-symmetrized for $\rho_{yx}$ data to eliminate the effect of electrode misalignment. For the analysis near QPT (Figs. 4d, 4e and 4f), an equivalent device width $W^*$ was used to accommodate different $L/W$ ratios among devices. Since scattering rates from sample boundaries are proportional to $1/(L \cdot W)$, which translates to $1/W^2$ for devices with $W \geq$ 5 μm and $1/(2W^2)$ for devices with $W \leq 2$ μm, $W^* = \sqrt{2}W$ was used to represent the equivalent width for devices with $W \leq 2$ μm for a meaningful comparison.

# Acknowledgements

We thank H. Liu for fruitful discussions. This work was supported by the National Natural Science Foundation of China (92065206, 11904053 and 92365201), the National Key Research and Development Program of China (2018YFA0307100) and the Innovation Program for Quantum Science and Technology (2021ZD0302502).

## Author contributions

Y.B., Y.L., Y.F., X.F. and K.H. conceived the experiment. Y.B. synthesized the samples, with the help of Y.C. Y.L. and J.L. performed electrical transport measurements. Y.L., Z.G., W.S. and Y.T. performed device fabrications. Y.B., Y.L., J.Q., C.-Z.C., H.J., X.C.X., Y.F., X.F., Q.-K.X. and K.H. analyzed and interpreted the transport data. Y.B., Y.F., X.F. and K.H. wrote the manuscript with input from all authors.

## Competing interests

The authors declare no competing interests.

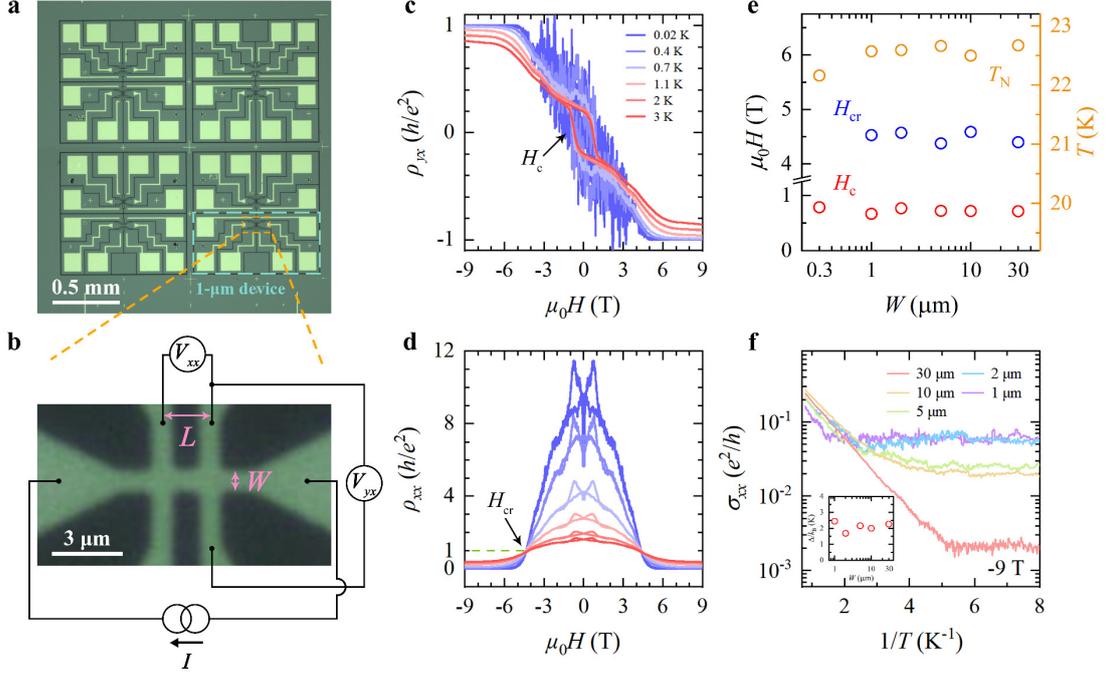

**Figure 1 | Characterizations of devices with different sizes based on the MBE-grown 5-SL MnBi$_2$Te$_4$ thin film. a**, The optical microscopic image of 8 devices fabricated in a 2 mm × 2 mm region without top gates. **b**, An enlarged image of the 2 μm ($L$) × 1 μm ($W$) device with schematic illustrations of the transport measurement setup. **c**, **d**, The $\rho_{yx}$–$\mu_0H$ and $\rho_{xx}$–$\mu_0H$ curves of the 30-μm device with $V_g$ fixed at the charge-neutral point at selected $T$, respectively. **e**, The $W$ dependences of $T_N$, $H_{cr}$ and $H_c$ (extracted at 2 K). **f**, The 1/$T$-dependent $\sigma_{xx}$ of each device at -9 T during the warming process. The inset shows the fitted gap values to thermal activation, $\sigma_{xx} \sim \exp(-\Delta/2k_BT)$, of each device.

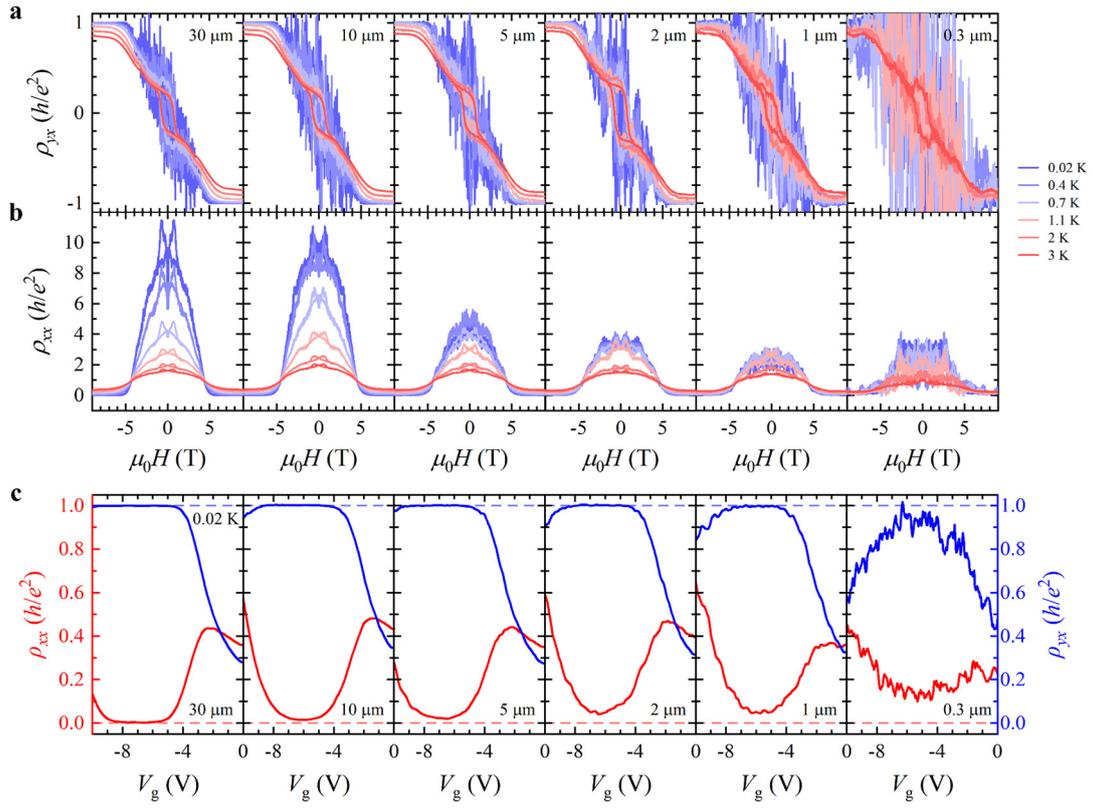

**Figure 2 | Transport results of devices with different sizes. a**, **b**, The temperature evolution of the $\rho_{yx}$–$\mu_0H$ and $\rho_{xx}$–$\mu_0H$ curves for each device. The $V_g$ of each device is fixed at the charge-neutral point during measurements. **c**, The $V_g$ dependences of $\rho_{yx}$ and $\rho_{xx}$ of each device at 20 mK and -9 T.

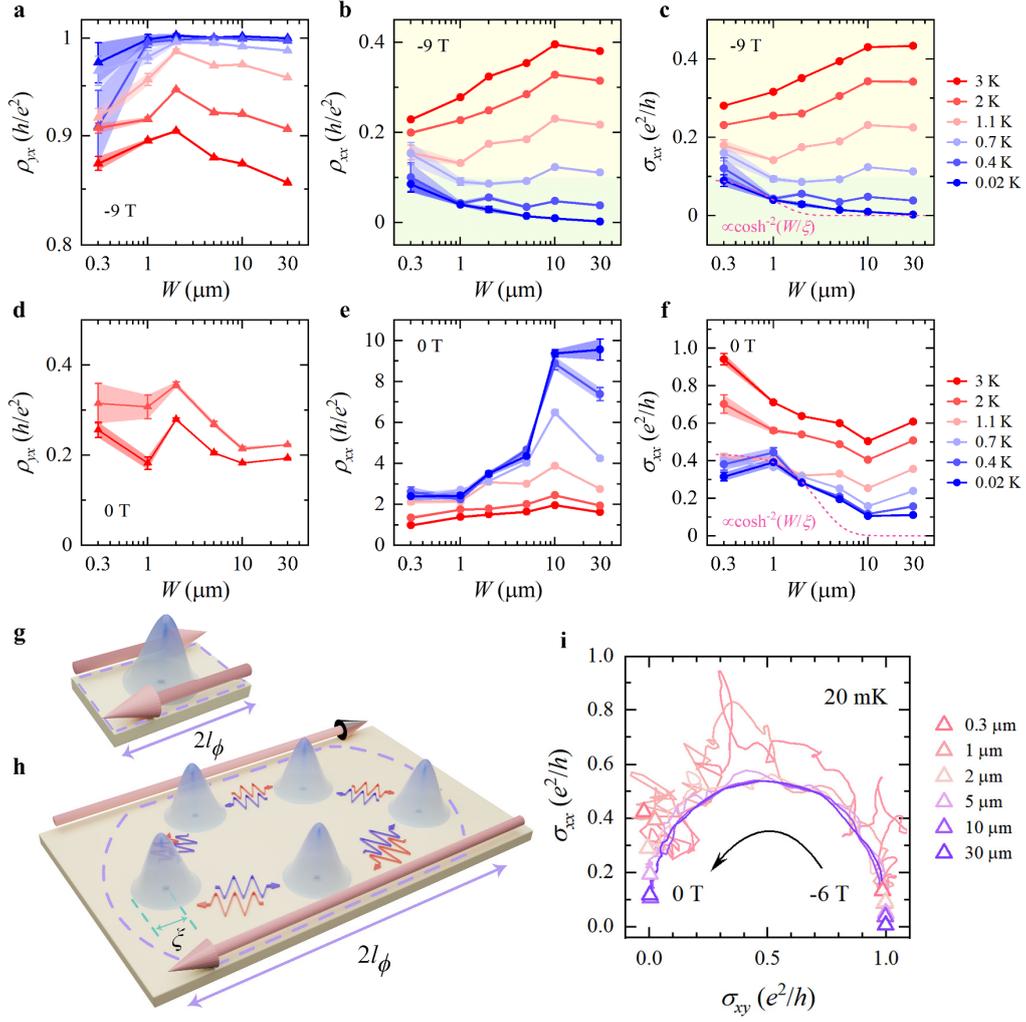

**Figure 3 | The interplay between localization and dephasing at two stable phases (-9 T and 0 T). a**, **b**, The $W$ dependences of $\rho_{yx}$ (-9 T) and $\rho_{xx}$ (-9 T) at selected $T$, respectively. **c**, The $W$-dependent $\sigma_{xx}$ (-9 T) at selected $T$. **d**, **e**, $\rho_{yx}$ (0 T)–$W$, $\rho_{xx}$ (0 T)–$W$ curves at selected $T$, respectively. Only the data of $\rho_{yx}$ (0 T) above 2 K are shown. **f**, The $W$-dependent $\sigma_{xx}$ (0 T) at selected $T$. The pink fitting curves in **c** and **f** are based on scattering matrix theory applied in quantum transport $\sigma_{xx} \propto \cosh^{-2}(W/\xi)$ with $\xi = 1$ μm and 3 μm, respectively. **g**, **h**, The scenarios of phase-coherent transport in the QAH state for small (**g**) and large (**h**) devices, respectively. In small devices, resonant tunneling between edge states at opposite edges through in-gap states can occur. In large devices, bulk electrons exhibit Anderson localization. **i**, The flow diagram of each device at 20 mK. Each curve illustrates the evolution of ($\sigma_{xy}$, $\sigma_{xx}$) from -6 T to 0 T, with endpoints marked as hollow triangles.

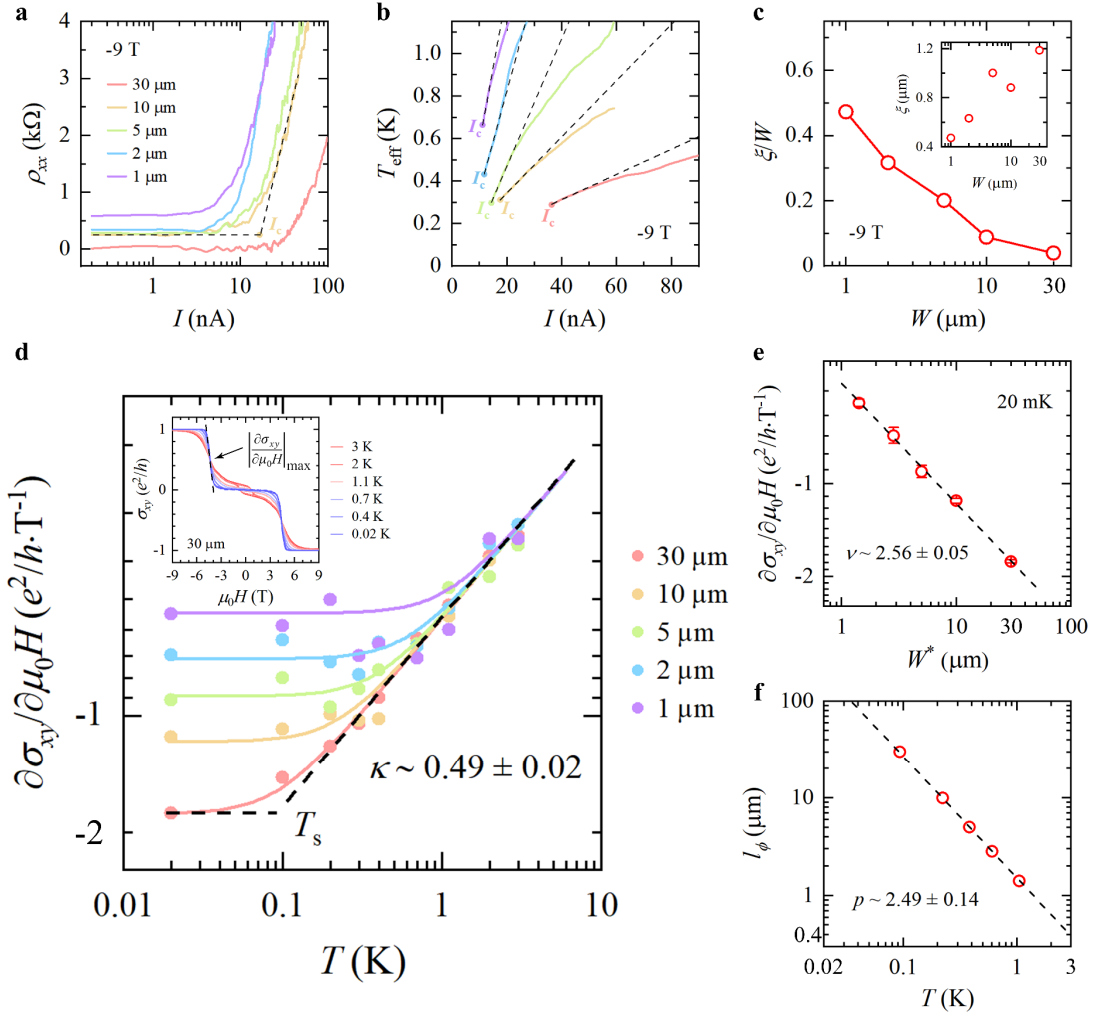

**Figure 4 | Estimations of the localization length and phase coherence length in the MnBi$_2$Te$_4$ thin film. a**, The $\rho_{xx}$–$I$ curve of each device measured at 20 mK and -9 T. $I$ represents the magnitude of the dc bias current. **b**, The $T_{eff}$–$I$ curve of each device at -9 T. **c**, The $W$ dependence of the renormalized $\xi/W$ under the QAH regime, exhibiting a localized behavior. The inset shows the estimated $\xi$ in each device at -9 T. All above measurements were conducted with $V_g$ fixed at the charge-neutral point. **d**, The double-logarithm plot of $\partial\sigma_{xy}/\partial\mu_0H$ and $T$ near $H_{cr}$. The colored curves represent fitting results as discussed in the main text. The dashed lines indicate the method used to define $T_s$. The inset shows the temperature evolution of $\sigma_{xy}$–$\mu_0H$ curves in the 30-μm device, where the dashed line with maximum $|\partial\sigma_{xy}/\partial\mu_0H|$ serves as a visual guide. **e**, The double-logarithm plot of $\partial\sigma_{xy}/\partial\mu_0H$ and equivalent device width $W^*$ at 20 mK. $W^*$ is adopted instead of $W$ to accommodate variations in the aspect ratios among devices (see Methods). **f**, The double-logarithm plot of $l_\phi$ and $T$ (that is, $W^*$ and $T_s$).